\documentclass[sn-mathphys, Numbered, iicol ]{sn-jnl} 

\usepackage{graphicx}%
\usepackage{multirow}%
\usepackage{amsmath,amssymb,amsfonts}%
\usepackage{amsthm}%
\usepackage{mathrsfs}%
\usepackage[title]{appendix}%
\usepackage{xcolor}%
\usepackage{textcomp}%
\usepackage{manyfoot}%
\usepackage{booktabs}%
\usepackage{algorithm}%
\usepackage{algorithmicx}%
\usepackage{algpseudocode}%
\usepackage{listings}%
\usepackage{upgreek}%
\usepackage{hyperref}

\begin{document}

\title{Superradiant and subradiant states in lifetime-limited organic molecules through laser-induced tuning}

\author[1]{\fnm{Christian} \sur{Lange}}

\author[2]{\fnm{Emma} \sur{Daggett}}

\author[1,2]{\fnm{Valentin} \sur{Walther}}

\author[2]{\fnm{Libai} \sur{Huang}} 

\author[1,2,*]{\fnm{Jonathan D } \sur{Hood}} 

\affil[1]{\orgdiv{Department of Physics and Astronomy}, \orgname{Purdue University}, \orgaddress{\city{West Lafayette}, \postcode{47907}, \state{IN}, \country{USA}}}  

\affil[2]{\orgdiv{Department of Chemistry}, \orgname{Purdue University}, \orgaddress{ \city{West Lafayette}, \postcode{47907}, \state{IN}, \country{USA}}}  

\affil[*]{Corresponding author: hoodjd@purdue.edu}

\abstract{
An array of radiatively coupled emitters is an exciting new platform for generating, storing, and manipulating quantum light. However, the simultaneous positioning and tuning of multiple lifetime-limited emitters into resonance remains a significant challenge.  Here we report the creation of superradiant and subradiant entangled states in pairs of lifetime-limited and sub-wavelength spaced organic molecules by permanently shifting them into resonance with laser-induced tuning.  The molecules are embedded as defects in an organic nanocrystal. The pump light redistributes charges in the nanocrystal and dramatically increases the likelihood of resonant molecules.  The frequency spectra, lifetimes, and second-order correlation agree with a simple quantum model.  This scalable tuning approach with organic molecules provides a pathway for observing collective quantum phenomena in sub-wavelength arrays of quantum emitters.  
}   


\maketitle

    \begin{figure*}[t!]%
    \centering
    \includegraphics[width=0.8\textwidth]{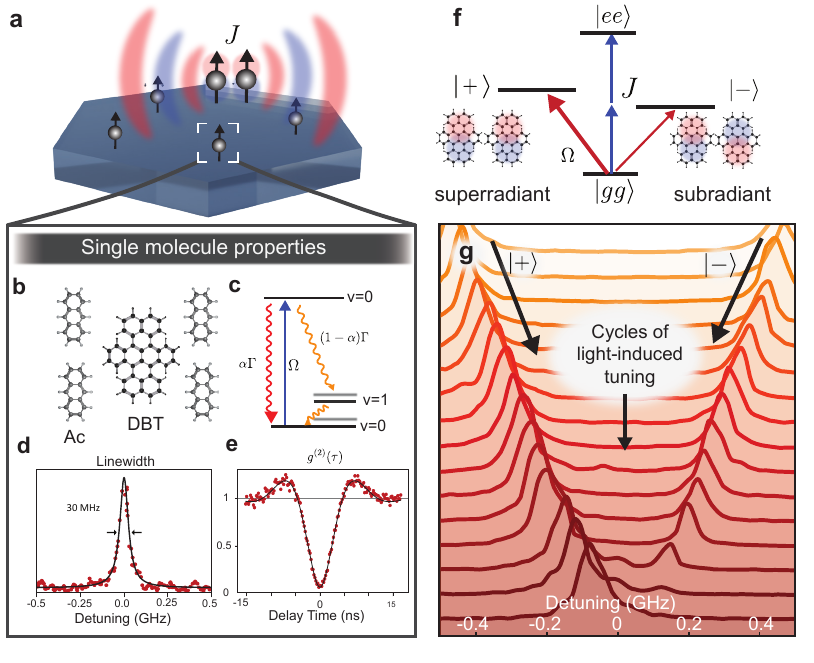}
    \caption{\textbf{Overview of DBT interactions} \textbf{(a)} Two DBT molecules in an anthracene crystal interact through their dipole moments with rate $J$.  \textbf{(b)}  The DBT molecule is embedded as a defect into an anthracene crystal and has a HOMO-LUMO transition near 785 nm with a linear dipole moment.  \textbf{(c)}  Level structure of DBT. The excited state decays with rate $\alpha \Gamma_0$ to the electronic and vibrational ground state, where $\alpha$ is the combined Debye-Waller/Franck-Condon factor. The excited state also decays with rate $(1-\alpha) \Gamma_0$ to higher vibrational states and their phonon sidebands.  The decay into the ground vibrational state is filtered out, and the decay to the vibrational levels is used for all fluorescence measurements. \textbf{(d)} Lifetime-limited linewidth scan of a single DBT molecule at 2.7 K.  \textbf{(e)} The $g^{(2)}(\tau)$ for a single molecule with $g^{(2)}(0)=0.065(9)$.  \textbf{(f)} This interaction leads to collective superradiant  $|+\rangle $ and subradiant $|- \rangle $ states. \textbf{(g)} Spectra of two molecules as they are tuned into resonance with intense illumination. The subradiant peak extinguishes as the detuning becomes smaller than the interaction $J$.  
    } \label{fig1}
    \end{figure*}


At sub-wavelength spacings, quantum emitters interact collectively with the electromagnetic field. Absorption and emission from separated atoms interfere, leading to super- and subradiant emitter states~\cite{dicke1954coherence, reitz2022cooperative}. The subradiant states form a decoherence free-subspace~\cite{lidar1998decoherence, facchinetti2016storing} useful for the manipulation of quantum states~\cite{beige2000driving} and the  simulation of many-body states~\cite{perczel2017topological, parmee2020signatures}. The collective emission also creates a wide variety of quantum states of light~\cite{porras2008collective, gonzalez2015deterministic, holzinger2020nanoscale}, which can be useful in quantum imaging and sensing~\cite{berchera2019quantum}. In regular arrays, collective emitters can function as light-matter interfaces of unit coupling, allowing the efficient storage and manipulation of light~\cite{bettles2016enhanced,facchinetti2016storing, asenjo2017exponential, rui2020subradiant, bekenstein2020quantum}. While optical superradiance has been observed in a wide variety of atomic~\cite{eschner2001light, goban2015superradiance, solano2017super, mcguyer2015precise, ferioli2021laser} and solid-state ensembles~\cite{hettich2002nanometer, scheibner2007superradiance, sipahigil2016integrated, blach2022superradiance}, the observation of individual states is hampered by three major requirements: sub-wavelength positioning, tuning into resonance, and minimizing dephasing from the environment. 
The realization of all three has only recently been demonstrated for two emitters using quantum dots coupled to a waveguide~\cite{tiranov2023collective} and organic molecules tuned by nano-electrodes~\cite{trebbia2022tailoring}, but the challenge remains to scale this up to large system sizes.

In this work, we demonstrate a new scalable method for tuning lifetime-limited organic molecules in close proximity into resonance. We observe the superradiant and subradiant states for a pair of molecules as they are brought into resonance using laser-induced tuning, as shown in Fig.~\ref{fig1}(g). Dibenzoterrylene molecules embedded as defects in an anthracene crystal~\cite{tamarat1999ten, toninelli2021single} are excellent quantum emitters and lifetime-limited below 3~K~\cite{clear2020phonon, wang2019turning}. The tuning light causes migration of charges within the nanocrystal~\cite{colautti2020laser}, locally decreasing inhomogeneous broadening. After tuning, the likelihood of pairs of molecules interacting is three orders of magnitude greater than would be expected for randomly distributed positions and frequencies. 

We characterize multiple molecule pairs at large detuning and near resonance and find excellent agreement of the linewidths, lifetimes, and second-order correlations with a master equation simulation.  The agreement with a simple model, ease of fabrication, and scalability of the tuning method demonstrate that organic molecules are a promising platform for creating cooperative phenomena in arrays of quantum emitters. 

\textbf{Methods:}  
Dibenzoterrylene (DBT) molecules, shown in Fig.~\ref{fig1}(b), are embedded as defects in anthracene nanocrystals~\cite{tamarat1999ten, pazzagli2018self, toninelli2021single}, with a HOMO-LUMO transition near 785 nm~\cite{nicolet2007single}. Below 3~K, the transition is lifetime-limited with high purity of single-photon emission, measured through linewidths and autocorrelation spectra, as shown in Fig.~\ref{fig1}(d, e). The excited state decays to the zero-phonon line of the ground vibrational state with a probability of 30\%, given by the product of the Franck-Condon and Debye-Waller factors. 

DBT and other polyaromatic hydrocarbons are leading candidates for solid-state quantum emitters and have demonstrated 94$\%$ indistinguishability for two photons from the same emitter~\cite{rezai2018coherence},  $70\%$ indistinguishability for photons from different emitters~\cite{duquennoy2023singular}, and a photon collection efficiency of 99$\%$~\cite{wrigge2008efficient}.

   \begin{figure*}[t!]%
    \centering
    \includegraphics[width=0.85\textwidth]{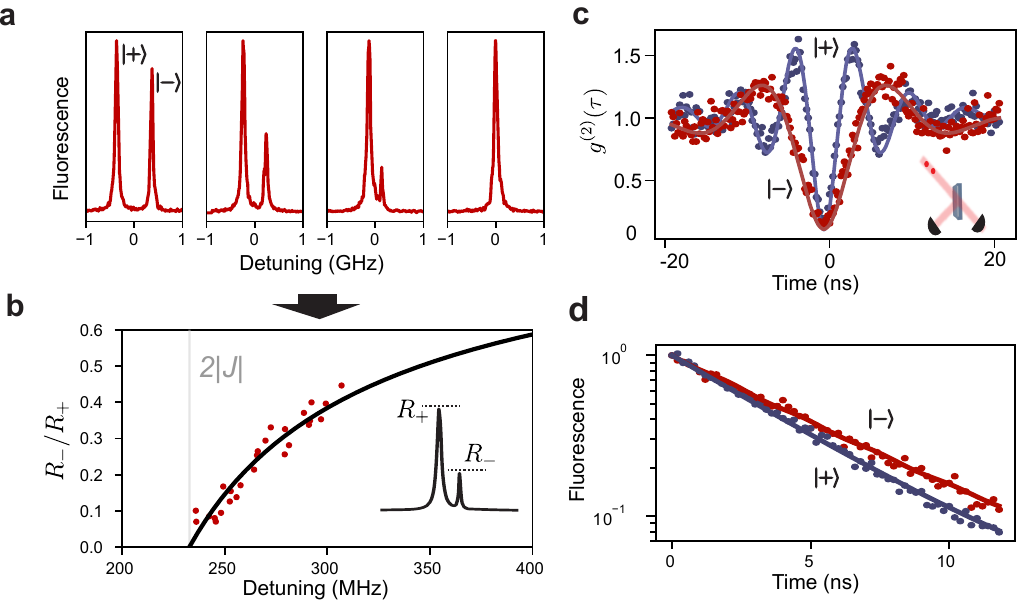}
    \caption{ \textbf{Spectroscopy of two molecules tuned into resonance.} \textbf{(a)}  Spectra of the superradiant $(\left|+\right>)$ and subradiant $(\left|-\right>)$ states of two coupled molecules as the molecules are tuned into resonance.  The tuning in $\textbf{(a)}$ was done with 170 cycles of 100-second exposure to $4 \, \mathrm{kW/cm^2}$ of 785 nm laser light.  \textbf{(b)} Ratio of the heights of the subradiant and superradiant peak as a function of detuning, fitted to a master equation simulation to obtain $J=-116 \, \mathrm{MHz}$. Spectra were obtained at an excitation intensity of $I=0.43 \, \mathrm{W / cm^2}$ \textbf{(c)} Second-order correlation functions $g^{(2)}(\tau)$  and \textbf{(d)}  lifetimes of a different pair of superradiant and subradiant peaks at $\Delta=2600 \, \mathrm{MHz}$ detuning. The superradiant (subradiant) state has an increased (decreased) Rabi-frequency, which results in faster (slower) oscillation of $g^{(2)}(\tau)$ for a given excitation power. The $g^{(2)}(\tau)$  functions are fitted to a master equation simulation to give $J=1020 \, \mathrm{MHz}$, $I/I_{\mathrm{sat}}=27$, and dephasing of 1 MHz. The lifetimes are fitted to give $\Gamma_0=33 \, \mathrm{MHz}$ and $\alpha=0.11$. }\label{fig2}
    \end{figure*}

Anthracene nanocrystals doped with a few hundred DBT molecules are self-assembled by precipitation from solution in a sonicator~\cite{pazzagli2018self} and vary from 200 nm to 1 $\upmu$m in size. The DBT molecule transition frequencies can be tuned by 100 GHz through exposure to intense laser light, as demonstrated in Ref.~\cite{colautti2020laser}.  The tuning persists after the pump light is turned off and is likely due to a photoionization process, whereby an electron from DBT is mobilized into the surrounding lattice, resulting in a Stark shift through the static electric field. We pump the nanocrystals with around 1 mW of 785~nm light focused through an objective to a waist of 1.5 $\upmu$m for tens of minutes.  We then look for a two-photon peak with a height that depends on power squared, which indicates an interaction as described below.  Out of twenty-five pumped nanocrystals, we observe ten signatures of interactions. 

The molecules interact through the electric fields of their oscillating dipole moments. The dipole moments are linear and all aligned to the crystalline lattice~\cite{nicolet2007single}. The Hamiltonian for a system of $N$ interacting dipoles with frequencies $\omega_i$ is
\begin{equation}
H =\sum_i^N \left( \omega_i - i \frac{\Gamma_0}{2} \right) \sigma^\dagger_i \sigma_i +  \sum_{i \ne j}^N \left(J_{ij} - i \frac{\Gamma_{ij}}{2} \right) \sigma^\dagger_i \sigma_j .
\end{equation}

The non-Hermitian rates $\Gamma_0$ and $\Gamma_{ij}$ are the spontaneous decay and dipole-dipole stimulated decay rates~\cite{reitz2022cooperative}. The coherent dipole-dipole interaction $J_{ij}$ generates energy exchange between molecules, and in the near-field is given by 
\begin{equation}
 J_{ij} = \frac{1}{4\pi\epsilon_0\epsilon_r}\frac{ \mathbf{d}_i\cdot \mathbf{d}_j - 3( \mathbf{d}_i\cdot \hat{r})( \mathbf{d}_j\cdot \hat{r})}{r^3}    
\end{equation}

As shown in the level structure in Fig.~\ref{fig1}(f), the singly-excited states  $|eg \rangle$ and $|ge \rangle$ are coupled by $J$ to form eigenstates $ |+ \rangle  = \sin{\theta} \, |eg \rangle + \cos{\theta} \, |ge \rangle $ and $|- \rangle  = \cos{\theta} \, |eg \rangle - \sin{\theta} \, |ge \rangle$ with detuning $ \tilde{\Delta}=  \sqrt{\Delta^2 + 4 J^2 }$. Here, $\tan{\theta} =( 2 J )/( \Delta + \tilde{\Delta} ) $.

As the pumping laser shifts interacting molecules into resonance, 
the anti-symmetric subradiant peak becomes dark to the symmetric drive Hamiltonian $H_\text{D} =  \sum_i \frac{\Omega_i}{2} (\sigma_i + \sigma_i^\dagger )$ and extinguishes when $\Delta < 2 J$.   When the molecules are fully entangled, the eigenstates are the purely symmetric and anti-symmetric states 
\begin{equation}
|\pm \rangle = (|eg \rangle \pm |ge \rangle )/\sqrt{2}.    
\end{equation}

In Fig.~\ref{fig2}(a), two coupled molecules are tuned into resonance, causing the subradiant state to narrow and decrease in intensity as it decouples from the symmetric laser probe.
After the extinction of the subradiant state, one pair of molecules remained resonant for over 24 hours of continuous monitoring. 
Other pairs of interacting molecules remained detuned by a few GHz despite undergoing tens of GHz of accumulated frequency shifting. 

 When the coupled states are degenerate, the magnitude of the coupling strength $J=-116$~MHz is equal to half the splitting of the eigenstates. The sign of the coupling strength depends on the relative orientation of the dipoles and determines which eigenstate has higher energy. In the H-aggregate (J-aggregate) orientation, the dipoles are perpendicular (parallel) to their separation vector, $J$ is positive (negative), and the superradiant state is higher (lower) in energy. 
 
 Fig.~\ref{fig2}(c-d) shows the $g^{(2)}(\tau)$ functions and lifetimes of a second pair of near-resonant molecules.  For a detuning of $\Delta = 2600$~MHz, the lifetimes were $\tau_+ =5.2$~ns and $\tau_- =4.3$~ns, whereas a non-interacting molecule typically has a lifetime of $\tau =4.5$~ns.  The $g^{(2)}(\tau)$ function is the probability of measuring two photons with time delay $\tau$.  The extracted Rabi frequencies are $\Omega_+=4.45\,\Gamma_0 $ and $\Omega_-=2.12 \,\Gamma_0$.  
 
The Rabi frequency exhibits a more pronounced effect because it is only related to the transition frequency with which the molecules interact.  The lifetime, in contrast, contains both the decay from the interacting transition and the decay to the other vibrational states and phonon sidebands.

    \begin{figure*}[t]
    \centering
    \includegraphics[width=0.7\textwidth]{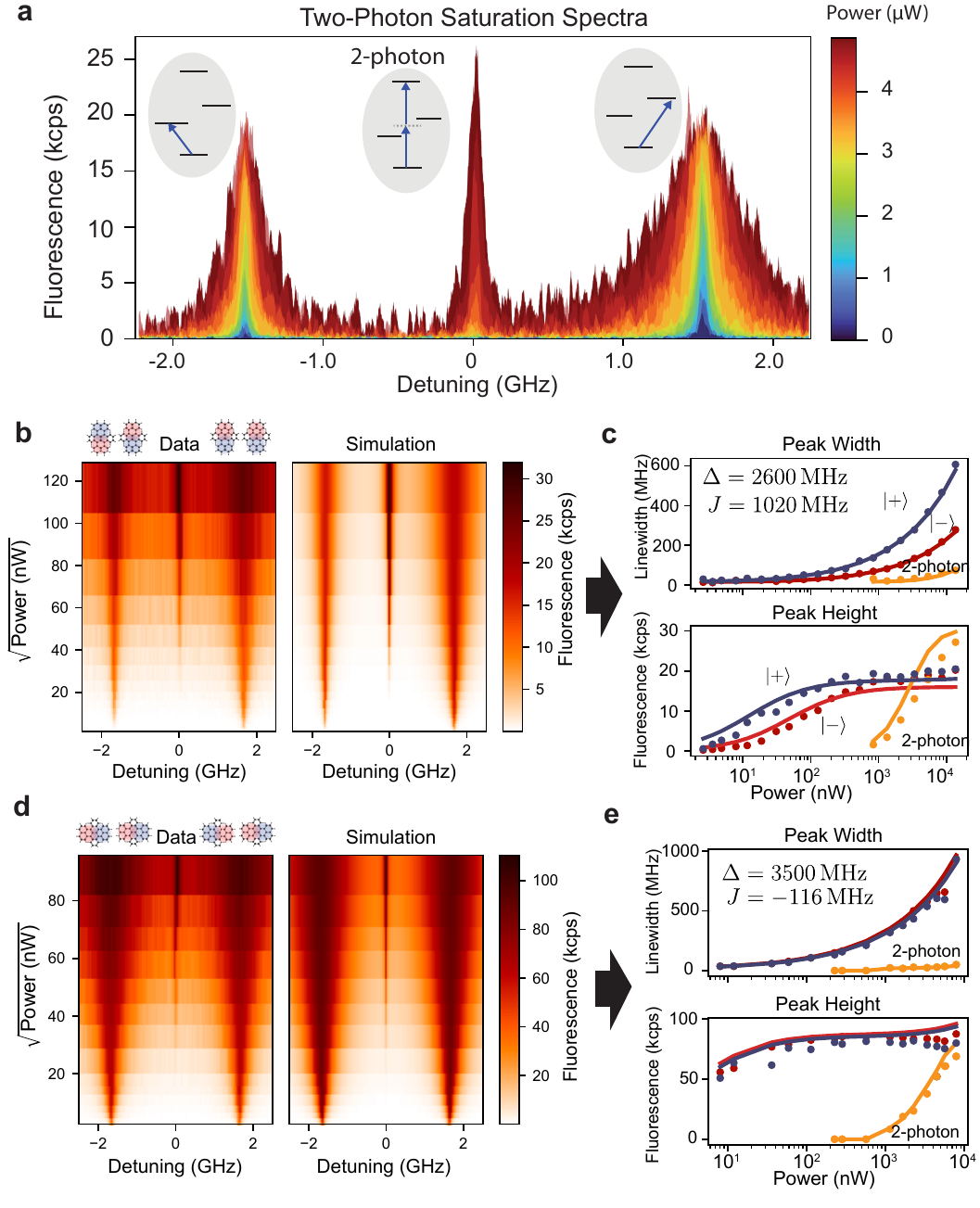} 
    \caption{\textbf{Simulation of saturation spectra}  \textbf{(a-c)} Saturation spectra for molecules in the H-aggregate orientation, where the dipoles are oriented perpendicular to the separation vector and the superradiant state is higher in energy than the subradiant. The spectra are aligned by the average of their superradiant and subradiant frequencies. As excitation power increases, the two-photon peak appears between the two dressed states. The heights and widths are extracted with Lorentzian fits and plotted in \textbf{(c)}. The simulation parameters are $\Gamma_0=33$ MHz, $J=1020$ MHz, $\alpha=0.11$, $\mathrm{dephasing}=1$ MHz, fitted through $g^{(2)}(\tau)$ and lifetime simulations. \textbf{(d-e)} The same simulation was performed for molecules in the J-aggregate orientation, with dipoles parallel to the separation vector and the superradiant state lower in energy than the subradiant state. The fit parameters are $\Gamma_0=37$  MHz, $J=-116$  MHz, $\alpha=0.135$, $\mathrm{dephasing}=1$  MHz. }\label{fig3}
    \end{figure*}

Fig.~\ref{fig3}(a) shows a series of fluorescence spectra of two interacting molecules with increasing excitation power. 
At high excitation power, the doubly-excited state $\left|ee\right>$ populates, and a two-photon peak emerges in the center of the coupled resonances. 
Fig.~\ref{fig3}(b-c) compares these spectra with a master equation simulation with $J=1020$~MHz, a DWFC factor $\alpha=0.11$, $1$~MHz dephasing, and $\Gamma_0=33$~MHz. These values are extracted from the $g^{(2)}(\tau)$ and lifetime measurements shown in Fig.~\ref{fig2}(c-d). 
To illustrate the agreement between experiment and theory, the scattering rates and linewidths of the three peaks are extracted with Lorentzian fits and plotted against the simulation values in Fig.~\ref{fig3}(c). 
Above saturation, the asymmetry of the superradiant and subradiant peaks depends on $J$ and $\alpha$. 
In contrast, the height of the two-photon peak is sensitive to $J$ and dephasing, and the saturated linewidths are primarily sensitive to $J$. 
Because the superradiant state is higher in energy than the subradiant, we can infer that the molecules are in the H-aggregate orientation, with dipoles roughly perpendicular to the separation vector. 
Fig.~\ref{fig3}(d-e) shows a similar set of data for a pair of molecules in the J-aggregate orientation, with $J=-116$ MHz, $\alpha=0.135$, $1$~MHz dephasing, and $\Gamma_0=37$~MHz, which are extracted from lifetime measurements and the extinction curve in Fig.~\ref{fig2}(b).
The FCDW factors of $\alpha=0.135$ and $0.11$ are lower than the reported value of 0.3 for a single DBT molecule.  This may be due to the close proximity of the molecules, differences in the synthesis, or misalignment of the dipoles.  

    \begin{figure*}[h!]%
    \centering
    \includegraphics[width=0.99\textwidth]{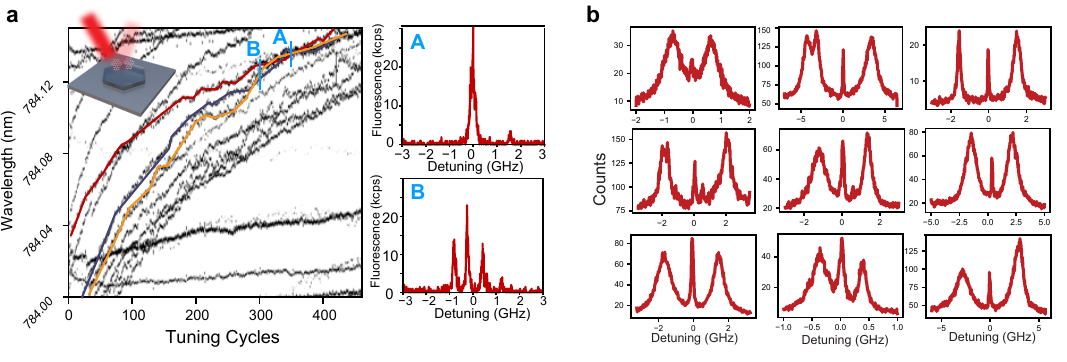}\caption{ \textbf{Laser-induced tuning} \textbf{(a)} Many molecules tuning in a single nanocrystal. The nanocrystal is pumped with $16 \, \mathrm{kW/cm^2}$ for 10 seconds between probes. Panels (1) and (2) show three resonances being tuned within a single linewidth.  The right two panels A and B are spectra showing the three molecules in and out of resonance.   \textbf{(b)} Nine two-photon peaks from separate nanocrystals, each pumped with 10 mW of light for 30 minutes. The coupled molecules exhibit both H-aggregate and J-aggregate orientations, as well as weak and strong interaction strength regimes. Several spectra indicate more than two interacting molecules. 
 }  \label{fig4}
    \end{figure*}

Fig.~\ref{fig4}(a) shows many molecules in a single nanocrystal shifting during successive pump cycles. The colored lines highlight three molecules being brought into resonance from a detuning of over 25 GHz, with spectra of the three molecules before and after shown to the right. Given the large inhomogeneous broadening, three molecules randomly shifting into resonance is unlikely, and this is an indicator of the tuning methods's ability to locally decrease inhomogeneous broadening. 

One likely explanation is that as the high-intensity light mobilizes charge carriers in the lattice, they migrate to minimize local electric field gradients contributing to inhomogeneous broadening. 
This picture could explain the consistency with which this effect can generate pairs of interacting molecules, as showcased in Fig.~\ref{fig4}(b). After laser-induced shifting, ten nanocrystals out of twenty-five had clear two-photon peaks. Interestingly, the coupled molecules commonly came into resonance with additional molecules, and some remained in resonance even while tuning. For example, the subradiant peak in the lower-right spectrum of Fig.~\ref{fig4}(b) is broader than the corresponding superradiant peak because it is overlapped with an additional resonance, as verified by the $g^{(2)}(\tau)$ and saturation spectra. This result is promising for the creation of many-body interactions and will be the subject of future research.

The length scale over which inhomogeneous broadening can be reduced is likely determined by the length scale of imperfections in the lattice, which cause inhomogeneous broadening. With an increase in the purity of nanocrystal synthesis or dopant density of high-purity synthesis methods like cosublimation, the length scale of this effect could be increased, allowing for the creation of many-body collective effects in the solid state. 

\textbf{Conclusion: } 
This work marks the first demonstration of permanently tuning lifetime-limited solid-state emitters into resonance and preparing the superradiant and subradiant states. 
Laser-induced tuning increases the likelihood of obtaining resonant molecules within a nanocrystal by three orders of magnitude, and two-photon peaks from interacting molecules are observed in $30\%$ of nanocrystals.   We characterized interacting molecules in different orientations, extracted the interaction strengths and branching ratios, and showed good agreement with a simple theory.  

The light-based tuning method can be scaled to larger arrays of emitters.  Molecules separated by more than 200 nm could also be addressed individually with light~\cite{colautti2020laser}.  The molecules could be positioned with sub-wavelength accuracy using techniques like nanoprinting~\cite{hail2019nanoprinting}. 
Organic molecules have also been shown to be compatible with nanophotonic devices~\cite{rattenbacher2019coherent}, which would enhance the Franck-Condon/Debye-Waller factor, create long-range interactions, and increase the photon  collection efficiency. 

Entangled states in collections of organic molecules could be used to generate, manipulate, and detect quantum light, measure electric fields with  enhanced sensitivity, or elucidate the role of dephasing and vibrational states in collective states~\cite{gurlek2021engineering, zirkelbach2022high}.






\textbf{Acknowledgments}
This work is supported by the NSF grant DMREF-2324299, and the US Department
of Energy (DOE) Office of Science through the Quantum Science Center
(QSC, a National Quantum Information Science Research Center). 
The authors thank Costanza Toninelli and Alex Clark for nanocrystal synthesis and characterization advice. 

\bibliography{bibliography}


\begin{thebibliography}{40}
\ifx \bisbn   \undefined \def \bisbn  #1{ISBN #1}\fi
\ifx \binits  \undefined \def \binits#1{#1}\fi
\ifx \bauthor  \undefined \def \bauthor#1{#1}\fi
\ifx \batitle  \undefined \def \batitle#1{#1}\fi
\ifx \bjtitle  \undefined \def \bjtitle#1{#1}\fi
\ifx \bvolume  \undefined \def \bvolume#1{\textbf{#1}}\fi
\ifx \byear  \undefined \def \byear#1{#1}\fi
\ifx \bissue  \undefined \def \bissue#1{#1}\fi
\ifx \bfpage  \undefined \def \bfpage#1{#1}\fi
\ifx \blpage  \undefined \def \blpage #1{#1}\fi
\ifx \burl  \undefined \def \burl#1{\textsf{#1}}\fi
\ifx \doiurl  \undefined \def \doiurl#1{\url{https://doi.org/#1}}\fi
\ifx \betal  \undefined \def \betal{\textit{et al.}}\fi
\ifx \binstitute  \undefined \def \binstitute#1{#1}\fi
\ifx \binstitutionaled  \undefined \def \binstitutionaled#1{#1}\fi
\ifx \bctitle  \undefined \def \bctitle#1{#1}\fi
\ifx \beditor  \undefined \def \beditor#1{#1}\fi
\ifx \bpublisher  \undefined \def \bpublisher#1{#1}\fi
\ifx \bbtitle  \undefined \def \bbtitle#1{#1}\fi
\ifx \bedition  \undefined \def \bedition#1{#1}\fi
\ifx \bseriesno  \undefined \def \bseriesno#1{#1}\fi
\ifx \blocation  \undefined \def \blocation#1{#1}\fi
\ifx \bsertitle  \undefined \def \bsertitle#1{#1}\fi
\ifx \bsnm \undefined \def \bsnm#1{#1}\fi
\ifx \bsuffix \undefined \def \bsuffix#1{#1}\fi
\ifx \bparticle \undefined \def \bparticle#1{#1}\fi
\ifx \barticle \undefined \def \barticle#1{#1}\fi
\bibcommenthead
\ifx \bconfdate \undefined \def \bconfdate #1{#1}\fi
\ifx \botherref \undefined \def \botherref #1{#1}\fi
\ifx \url \undefined \def \url#1{\textsf{#1}}\fi
\ifx \bchapter \undefined \def \bchapter#1{#1}\fi
\ifx \bbook \undefined \def \bbook#1{#1}\fi
\ifx \bcomment \undefined \def \bcomment#1{#1}\fi
\ifx \oauthor \undefined \def \oauthor#1{#1}\fi
\ifx \citeauthoryear \undefined \def \citeauthoryear#1{#1}\fi
\ifx \endbibitem  \undefined \def \endbibitem {}\fi
\ifx \bconflocation  \undefined \def \bconflocation#1{#1}\fi
\ifx \arxivurl  \undefined \def \arxivurl#1{\textsf{#1}}\fi
\csname PreBibitemsHook\endcsname

\bibitem[\protect\citeauthoryear{Dicke}{1954}]{dicke1954coherence}
\begin{barticle}
\bauthor{\bsnm{Dicke}, \binits{R.H.}}:
\batitle{Coherence in spontaneous radiation processes}.
\bjtitle{Phys. Rev.}
\bvolume{93},
\bfpage{99}--\blpage{110}
(\byear{1954})
\doiurl{10.1103/PhysRev.93.99}
\end{barticle}
\endbibitem

\bibitem[\protect\citeauthoryear{Reitz et~al.}{2022}]{reitz2022cooperative}
\begin{barticle}
\bauthor{\bsnm{Reitz}, \binits{M.}},
\bauthor{\bsnm{Sommer}, \binits{C.}},
\bauthor{\bsnm{Genes}, \binits{C.}}:
\batitle{Cooperative quantum phenomena in light-matter platforms}.
\bjtitle{PRX Quantum}
\bvolume{3},
\bfpage{010201}
(\byear{2022})
\doiurl{10.1103/PRXQuantum.3.010201}
\end{barticle}
\endbibitem

\bibitem[\protect\citeauthoryear{Lidar et~al.}{1998}]{lidar1998decoherence}
\begin{barticle}
\bauthor{\bsnm{Lidar}, \binits{D.A.}},
\bauthor{\bsnm{Chuang}, \binits{I.L.}},
\bauthor{\bsnm{Whaley}, \binits{K.B.}}:
\batitle{Decoherence-free subspaces for quantum computation}.
\bjtitle{Phys. Rev. Lett.}
\bvolume{81},
\bfpage{2594}--\blpage{2597}
(\byear{1998})
\doiurl{10.1103/PhysRevLett.81.2594}
\end{barticle}
\endbibitem

\bibitem[\protect\citeauthoryear{Facchinetti
  et~al.}{2016}]{facchinetti2016storing}
\begin{barticle}
\bauthor{\bsnm{Facchinetti}, \binits{G.}},
\bauthor{\bsnm{Jenkins}, \binits{S.D.}},
\bauthor{\bsnm{Ruostekoski}, \binits{J.}}:
\batitle{Storing light with subradiant correlations in arrays of atoms}.
\bjtitle{t.}
\bvolume{117},
\bfpage{243601}
(\byear{2016})
\doiurl{10.1103/PhysRevLett.117.243601}
\end{barticle}
\endbibitem

\bibitem[\protect\citeauthoryear{Beige et~al.}{2000}]{beige2000driving}
\begin{barticle}
\bauthor{\bsnm{Beige}, \binits{A.}},
\bauthor{\bsnm{Braun}, \binits{D.}},
\bauthor{\bsnm{Knight}, \binits{P.L.}}:
\batitle{Driving atoms into decoherence-free states}.
\bjtitle{New Journal of Physics}
\bvolume{2}(\bissue{1}),
\bfpage{22}
(\byear{2000})
\doiurl{10.1088/1367-2630/2/1/322}
\end{barticle}
\endbibitem

\bibitem[\protect\citeauthoryear{Perczel et~al.}{2017}]{perczel2017topological}
\begin{barticle}
\bauthor{\bsnm{Perczel}, \binits{J.}},
\bauthor{\bsnm{Borregaard}, \binits{J.}},
\bauthor{\bsnm{Chang}, \binits{D.E.}},
\bauthor{\bsnm{Pichler}, \binits{H.}},
\bauthor{\bsnm{Yelin}, \binits{S.F.}},
\bauthor{\bsnm{Zoller}, \binits{P.}},
\bauthor{\bsnm{Lukin}, \binits{M.D.}}:
\batitle{Topological quantum optics in two-dimensional atomic arrays}.
\bjtitle{Phys. Rev. Lett.}
\bvolume{119},
\bfpage{023603}
(\byear{2017})
\doiurl{10.1103/PhysRevLett.119.023603}
\end{barticle}
\endbibitem

\bibitem[\protect\citeauthoryear{Parmee and
  Ruostekoski}{2020}]{parmee2020signatures}
\begin{barticle}
\bauthor{\bsnm{Parmee}, \binits{C.D.}},
\bauthor{\bsnm{Ruostekoski}, \binits{J.}}:
\batitle{Signatures of optical phase transitions in superradiant and subradiant
  atomic arrays}.
\bjtitle{Commun. Phys.}
\bvolume{3}(\bissue{1}),
\bfpage{205}
(\byear{2020})
\doiurl{10.1038/s42005-020-00476-1}
\end{barticle}
\endbibitem

\bibitem[\protect\citeauthoryear{Porras and Cirac}{2008}]{porras2008collective}
\begin{barticle}
\bauthor{\bsnm{Porras}, \binits{D.}},
\bauthor{\bsnm{Cirac}, \binits{J.I.}}:
\batitle{Collective generation of quantum states of light by entangled atoms}.
\bjtitle{Phys. Rev. A}
\bvolume{78},
\bfpage{053816}
(\byear{2008})
\doiurl{10.1103/PhysRevA.78.053816}
\end{barticle}
\endbibitem

\bibitem[\protect\citeauthoryear{Gonz\'alez-Tudela
  et~al.}{2015}]{gonzalez2015deterministic}
\begin{barticle}
\bauthor{\bsnm{Gonz\'alez-Tudela}, \binits{A.}},
\bauthor{\bsnm{Paulisch}, \binits{V.}},
\bauthor{\bsnm{Chang}, \binits{D.E.}},
\bauthor{\bsnm{Kimble}, \binits{H.J.}},
\bauthor{\bsnm{Cirac}, \binits{J.I.}}:
\batitle{Deterministic generation of arbitrary photonic states assisted by
  dissipation}.
\bjtitle{Phys. Rev. Lett.}
\bvolume{115},
\bfpage{163603}
(\byear{2015})
\doiurl{10.1103/PhysRevLett.115.163603}
\end{barticle}
\endbibitem

\bibitem[\protect\citeauthoryear{Holzinger
  et~al.}{2020}]{holzinger2020nanoscale}
\begin{barticle}
\bauthor{\bsnm{Holzinger}, \binits{R.}},
\bauthor{\bsnm{Plankensteiner}, \binits{D.}},
\bauthor{\bsnm{Ostermann}, \binits{L.}},
\bauthor{\bsnm{Ritsch}, \binits{H.}}:
\batitle{Nanoscale coherent light source}.
\bjtitle{Phys. Rev. Lett.}
\bvolume{124},
\bfpage{253603}
(\byear{2020})
\doiurl{10.1103/PhysRevLett.124.253603}
\end{barticle}
\endbibitem

\bibitem[\protect\citeauthoryear{Berchera and
  Degiovanni}{2019}]{berchera2019quantum}
\begin{barticle}
\bauthor{\bsnm{Berchera}, \binits{I.R.}},
\bauthor{\bsnm{Degiovanni}, \binits{I.P.}}:
\batitle{Quantum imaging with sub-poissonian light: challenges and perspectives
  in optical metrology}.
\bjtitle{Metrologia}
\bvolume{56}(\bissue{2}),
\bfpage{024001}
(\byear{2019})
\doiurl{10.1088/1681-7575/aaf7b2}
\end{barticle}
\endbibitem

\bibitem[\protect\citeauthoryear{Bettles et~al.}{2016}]{bettles2016enhanced}
\begin{barticle}
\bauthor{\bsnm{Bettles}, \binits{R.J.}},
\bauthor{\bsnm{Gardiner}, \binits{S.A.}},
\bauthor{\bsnm{Adams}, \binits{C.S.}}:
\batitle{Enhanced optical cross section via collective coupling of atomic
  dipoles in a 2d array}.
\bjtitle{Phys. Rev. Lett.}
\bvolume{116},
\bfpage{103602}
(\byear{2016})
\doiurl{10.1103/PhysRevLett.116.103602}
\end{barticle}
\endbibitem

\bibitem[\protect\citeauthoryear{Asenjo-Garcia
  et~al.}{2017}]{asenjo2017exponential}
\begin{barticle}
\bauthor{\bsnm{Asenjo-Garcia}, \binits{A.}},
\bauthor{\bsnm{Moreno-Cardoner}, \binits{M.}},
\bauthor{\bsnm{Albrecht}, \binits{A.}},
\bauthor{\bsnm{Kimble}, \binits{H.J.}},
\bauthor{\bsnm{Chang}, \binits{D.E.}}:
\batitle{Exponential improvement in photon storage fidelities using subradiance
  and ``selective radiance'' in atomic arrays}.
\bjtitle{Phys. Rev. X}
\bvolume{7},
\bfpage{031024}
(\byear{2017})
\doiurl{10.1103/PhysRevX.7.031024}
\end{barticle}
\endbibitem

\bibitem[\protect\citeauthoryear{Rui et~al.}{2020}]{rui2020subradiant}
\begin{barticle}
\bauthor{\bsnm{Rui}, \binits{J.}},
\bauthor{\bsnm{Wei}, \binits{D.}},
\bauthor{\bsnm{Rubio-Abadal}, \binits{A.}},
\bauthor{\bsnm{Hollerith}, \binits{S.}},
\bauthor{\bsnm{Zeiher}, \binits{J.}},
\bauthor{\bsnm{Stamper-Kurn}, \binits{D.M.}},
\bauthor{\bsnm{Gross}, \binits{C.}},
\bauthor{\bsnm{Bloch}, \binits{I.}}:
\batitle{A subradiant optical mirror formed by a single structured atomic
  layer}.
\bjtitle{Nature}
\bvolume{583}(\bissue{7816}),
\bfpage{369}--\blpage{374}
(\byear{2020})
\doiurl{10.1038/s41586-020-2463-x}
\end{barticle}
\endbibitem

\bibitem[\protect\citeauthoryear{Bekenstein
  et~al.}{2020}]{bekenstein2020quantum}
\begin{barticle}
\bauthor{\bsnm{Bekenstein}, \binits{R.}},
\bauthor{\bsnm{Pikovski}, \binits{I.}},
\bauthor{\bsnm{Pichler}, \binits{H.}},
\bauthor{\bsnm{Shahmoon}, \binits{E.}},
\bauthor{\bsnm{Yelin}, \binits{S.F.}},
\bauthor{\bsnm{Lukin}, \binits{M.D.}}:
\batitle{Quantum metasurfaces with atom arrays}.
\bjtitle{Nat. Phys.}
\bvolume{16}(\bissue{6}),
\bfpage{676}--\blpage{681}
(\byear{2020})
\doiurl{10.1038/s41567-020-0845-5}
\end{barticle}
\endbibitem

\bibitem[\protect\citeauthoryear{Eschner et~al.}{2001}]{eschner2001light}
\begin{barticle}
\bauthor{\bsnm{Eschner}, \binits{J.}},
\bauthor{\bsnm{Raab}, \binits{C.}},
\bauthor{\bsnm{Schmidt-Kaler}, \binits{F.}},
\bauthor{\bsnm{Blatt}, \binits{R.}}:
\batitle{Light interference from single atoms and their mirror images}.
\bjtitle{Nature}
\bvolume{413}(\bissue{6855}),
\bfpage{495}--\blpage{498}
(\byear{2001})
\doiurl{10.1038/35097017}
\end{barticle}
\endbibitem

\bibitem[\protect\citeauthoryear{Goban et~al.}{2015}]{goban2015superradiance}
\begin{barticle}
\bauthor{\bsnm{Goban}, \binits{A.}},
\bauthor{\bsnm{Hung}, \binits{C.-L.}},
\bauthor{\bsnm{Hood}, \binits{J.D.}},
\bauthor{\bsnm{Yu}, \binits{S.-P.}},
\bauthor{\bsnm{Muniz}, \binits{J.A.}},
\bauthor{\bsnm{Painter}, \binits{O.}},
\bauthor{\bsnm{Kimble}, \binits{H.J.}}:
\batitle{Superradiance for atoms trapped along a photonic crystal waveguide}.
\bjtitle{Phys. Rev. Lett.}
\bvolume{115},
\bfpage{063601}
(\byear{2015})
\doiurl{10.1103/PhysRevLett.115.063601}
\end{barticle}
\endbibitem

\bibitem[\protect\citeauthoryear{Solano et~al.}{2017}]{solano2017super}
\begin{barticle}
\bauthor{\bsnm{Solano}, \binits{P.}},
\bauthor{\bsnm{Barberis-Blostein}, \binits{P.}},
\bauthor{\bsnm{Fatemi}, \binits{F.K.}},
\bauthor{\bsnm{Orozco}, \binits{L.A.}},
\bauthor{\bsnm{Rolston}, \binits{S.L.}}:
\batitle{Super-radiance reveals infinite-range dipole interactions through a
  nanofiber}.
\bjtitle{Nat. Commun.}
\bvolume{8}(\bissue{1}),
\bfpage{1857}
(\byear{2017})
\doiurl{10.1038/s41467-017-01994-3}
\end{barticle}
\endbibitem

\bibitem[\protect\citeauthoryear{McGuyer et~al.}{2015}]{mcguyer2015precise}
\begin{barticle}
\bauthor{\bsnm{McGuyer}, \binits{B.H.}},
\bauthor{\bsnm{McDonald}, \binits{M.}},
\bauthor{\bsnm{Iwata}, \binits{G.Z.}},
\bauthor{\bsnm{Tarallo}, \binits{M.G.}},
\bauthor{\bsnm{Skomorowski}, \binits{W.}},
\bauthor{\bsnm{Moszynski}, \binits{R.}},
\bauthor{\bsnm{Zelevinsky}, \binits{T.}}:
\batitle{Precise study of asymptotic physics with subradiant ultracold
  molecules}.
\bjtitle{Nat. Phys.}
\bvolume{11}(\bissue{1}),
\bfpage{32}--\blpage{36}
(\byear{2015})
\doiurl{10.1038/nphys3182}
\end{barticle}
\endbibitem

\bibitem[\protect\citeauthoryear{Ferioli et~al.}{2021}]{ferioli2021laser}
\begin{barticle}
\bauthor{\bsnm{Ferioli}, \binits{G.}},
\bauthor{\bsnm{Glicenstein}, \binits{A.}},
\bauthor{\bsnm{Robicheaux}, \binits{F.}},
\bauthor{\bsnm{Sutherland}, \binits{R.}},
\bauthor{\bsnm{Browaeys}, \binits{A.}},
\bauthor{\bsnm{Ferrier-Barbut}, \binits{I.}}:
\batitle{Laser-driven superradiant ensembles of two-level atoms near dicke
  regime}.
\bjtitle{Phys. Rev. Lett.}
\bvolume{127}(\bissue{24}),
\bfpage{243602}
(\byear{2021})
\end{barticle}
\endbibitem

\bibitem[\protect\citeauthoryear{Hettich et~al.}{2002}]{hettich2002nanometer}
\begin{barticle}
\bauthor{\bsnm{Hettich}, \binits{C.}},
\bauthor{\bsnm{Schmitt}, \binits{C.}},
\bauthor{\bsnm{Zitzmann}, \binits{J.}},
\bauthor{\bsnm{Kuhn}, \binits{S.}},
\bauthor{\bsnm{Gerhardt}, \binits{I.}},
\bauthor{\bsnm{Sandoghdar}, \binits{V.}}:
\batitle{Nanometer resolution and coherent optical dipole coupling of two
  individual molecules}.
\bjtitle{Science}
\bvolume{298}(\bissue{5592}),
\bfpage{385}--\blpage{389}
(\byear{2002})
\doiurl{10.1126/science.1075606}
\end{barticle}
\endbibitem

\bibitem[\protect\citeauthoryear{Scheibner
  et~al.}{2007}]{scheibner2007superradiance}
\begin{barticle}
\bauthor{\bsnm{Scheibner}, \binits{M.}},
\bauthor{\bsnm{Schmidt}, \binits{T.}},
\bauthor{\bsnm{Worschech}, \binits{L.}},
\bauthor{\bsnm{Forchel}, \binits{A.}},
\bauthor{\bsnm{Bacher}, \binits{G.}},
\bauthor{\bsnm{Passow}, \binits{T.}},
\bauthor{\bsnm{Hommel}, \binits{D.}}:
\batitle{Superradiance of quantum dots}.
\bjtitle{Nat. Phys.}
\bvolume{3}(\bissue{2}),
\bfpage{106}--\blpage{110}
(\byear{2007})
\doiurl{10.1038/nphys494}
\end{barticle}
\endbibitem

\bibitem[\protect\citeauthoryear{Sipahigil
  et~al.}{2016}]{sipahigil2016integrated}
\begin{barticle}
\bauthor{\bsnm{Sipahigil}, \binits{A.}},
\bauthor{\bsnm{Evans}, \binits{R.E.}},
\bauthor{\bsnm{Sukachev}, \binits{D.D.}},
\bauthor{\bsnm{Burek}, \binits{M.J.}},
\bauthor{\bsnm{Borregaard}, \binits{J.}},
\bauthor{\bsnm{Bhaskar}, \binits{M.K.}},
\bauthor{\bsnm{Nguyen}, \binits{C.T.}},
\bauthor{\bsnm{Pacheco}, \binits{J.L.}},
\bauthor{\bsnm{Atikian}, \binits{H.A.}},
\bauthor{\bsnm{Meuwly}, \binits{C.}},
\bauthor{\bsnm{Camacho}, \binits{R.M.}},
\bauthor{\bsnm{Jelezko}, \binits{F.}},
\bauthor{\bsnm{Bielejec}, \binits{E.}},
\bauthor{\bsnm{Park}, \binits{H.}},
\bauthor{\bsnm{Lončar}, \binits{M.}},
\bauthor{\bsnm{Lukin}, \binits{M.D.}}:
\batitle{An integrated diamond nanophotonics platform for quantum-optical
  networks}.
\bjtitle{Science}
\bvolume{354}(\bissue{6314}),
\bfpage{847}--\blpage{850}
(\byear{2016})
\doiurl{10.1126/science.aah6875}
\end{barticle}
\endbibitem

\bibitem[\protect\citeauthoryear{Blach et~al.}{2022}]{blach2022superradiance}
\begin{barticle}
\bauthor{\bsnm{Blach}, \binits{D.D.}},
\bauthor{\bsnm{Lumsargis}, \binits{V.A.}},
\bauthor{\bsnm{Clark}, \binits{D.E.}},
\bauthor{\bsnm{Chuang}, \binits{C.}},
\bauthor{\bsnm{Wang}, \binits{K.}},
\bauthor{\bsnm{Dou}, \binits{L.}},
\bauthor{\bsnm{Schaller}, \binits{R.D.}},
\bauthor{\bsnm{Cao}, \binits{J.}},
\bauthor{\bsnm{Li}, \binits{C.W.}},
\bauthor{\bsnm{Huang}, \binits{L.}}:
\batitle{Superradiance and exciton delocalization in perovskite quantum dot
  superlattices}.
\bjtitle{Nano Lett.}
\bvolume{22}(\bissue{19}),
\bfpage{7811}--\blpage{7818}
(\byear{2022})
\doiurl{10.1021/acs.nanolett.2c02427}
\end{barticle}
\endbibitem

\bibitem[\protect\citeauthoryear{Tiranov et~al.}{2023}]{tiranov2023collective}
\begin{barticle}
\bauthor{\bsnm{Tiranov}, \binits{A.}},
\bauthor{\bsnm{Angelopoulou}, \binits{V.}},
\bauthor{\bsnm{Diepen}, \binits{C.J.}},
\bauthor{\bsnm{Schrinski}, \binits{B.}},
\bauthor{\bsnm{Sandberg}, \binits{O.A.D.}},
\bauthor{\bsnm{Wang}, \binits{Y.}},
\bauthor{\bsnm{Midolo}, \binits{L.}},
\bauthor{\bsnm{Scholz}, \binits{S.}},
\bauthor{\bsnm{Wieck}, \binits{A.D.}},
\bauthor{\bsnm{Ludwig}, \binits{A.}},
\bauthor{\bsnm{Sørensen}, \binits{A.S.}},
\bauthor{\bsnm{Lodahl}, \binits{P.}}:
\batitle{Collective super- and subradiant dynamics between distant optical
  quantum emitters}.
\bjtitle{Science}
\bvolume{379}(\bissue{6630}),
\bfpage{389}--\blpage{393}
(\byear{2023})
\doiurl{10.1126/science.ade9324}
\end{barticle}
\endbibitem

\bibitem[\protect\citeauthoryear{Trebbia et~al.}{2022}]{trebbia2022tailoring}
\begin{barticle}
\bauthor{\bsnm{Trebbia}, \binits{J.-B.}},
\bauthor{\bsnm{Deplano}, \binits{Q.}},
\bauthor{\bsnm{Tamarat}, \binits{P.}},
\bauthor{\bsnm{Lounis}, \binits{B.}}:
\batitle{Tailoring the superradiant and subradiant nature of two coherently
  coupled quantum emitters}.
\bjtitle{Nat. Commun.}
\bvolume{13}(\bissue{1}),
\bfpage{2962}
(\byear{2022})
\doiurl{10.1038/s41467-022-30672-2}
\end{barticle}
\endbibitem

\bibitem[\protect\citeauthoryear{Tamarat et~al.}{2000}]{tamarat1999ten}
\begin{barticle}
\bauthor{\bsnm{Tamarat}, \binits{P.}},
\bauthor{\bsnm{Maali}, \binits{A.}},
\bauthor{\bsnm{Lounis}, \binits{B.}},
\bauthor{\bsnm{Orrit}, \binits{M.}}:
\batitle{Ten years of single-molecule spectroscopy}.
\bjtitle{The Journal of Physical Chemistry A}
\bvolume{104}(\bissue{1}),
\bfpage{1}--\blpage{16}
(\byear{2000})
\doiurl{10.1021/jp992505l}
\end{barticle}
\endbibitem

\bibitem[\protect\citeauthoryear{Toninelli et~al.}{2021}]{toninelli2021single}
\begin{barticle}
\bauthor{\bsnm{Toninelli}, \binits{C.}},
\bauthor{\bsnm{Gerhardt}, \binits{I.}},
\bauthor{\bsnm{Clark}, \binits{A.}},
\bauthor{\bsnm{Reserbat-Plantey}, \binits{A.}},
\bauthor{\bsnm{G{\"o}tzinger}, \binits{S.}},
\bauthor{\bsnm{Ristanovi{\'c}}, \binits{Z.}},
\bauthor{\bsnm{Colautti}, \binits{M.}},
\bauthor{\bsnm{Lombardi}, \binits{P.}},
\bauthor{\bsnm{Major}, \binits{K.}},
\bauthor{\bsnm{Deperasi{\'n}ska}, \binits{I.}}, \betal:
\batitle{Single organic molecules for photonic quantum technologies}.
\bjtitle{Nat. Mater.}
\bvolume{20}(\bissue{12}),
\bfpage{1615}--\blpage{1628}
(\byear{2021})
\doiurl{10.1038/s41563-021-00987-4}
\end{barticle}
\endbibitem

\bibitem[\protect\citeauthoryear{Clear et~al.}{2020}]{clear2020phonon}
\begin{barticle}
\bauthor{\bsnm{Clear}, \binits{C.}},
\bauthor{\bsnm{Schofield}, \binits{R.C.}},
\bauthor{\bsnm{Major}, \binits{K.D.}},
\bauthor{\bsnm{Iles-Smith}, \binits{J.}},
\bauthor{\bsnm{Clark}, \binits{A.S.}},
\bauthor{\bsnm{McCutcheon}, \binits{D.P.S.}}:
\batitle{Phonon-induced optical dephasing in single organic molecules}.
\bjtitle{Phys. Rev. Lett.}
\bvolume{124},
\bfpage{153602}
(\byear{2020})
\doiurl{10.1103/PhysRevLett.124.153602}
\end{barticle}
\endbibitem

\bibitem[\protect\citeauthoryear{Wang et~al.}{2019}]{wang2019turning}
\begin{barticle}
\bauthor{\bsnm{Wang}, \binits{D.}},
\bauthor{\bsnm{Kelkar}, \binits{H.}},
\bauthor{\bsnm{Martin-Cano}, \binits{D.}},
\bauthor{\bsnm{Rattenbacher}, \binits{D.}},
\bauthor{\bsnm{Shkarin}, \binits{A.}},
\bauthor{\bsnm{Utikal}, \binits{T.}},
\bauthor{\bsnm{G{\"o}tzinger}, \binits{S.}},
\bauthor{\bsnm{Sandoghdar}, \binits{V.}}:
\batitle{Turning a molecule into a coherent two-level quantum system}.
\bjtitle{Nat. Phys.}
\bvolume{15}(\bissue{5}),
\bfpage{483}--\blpage{489}
(\byear{2019})
\doiurl{10.1038/s41567-019-0436-5}
\end{barticle}
\endbibitem

\bibitem[\protect\citeauthoryear{Colautti et~al.}{2020}]{colautti2020laser}
\begin{barticle}
\bauthor{\bsnm{Colautti}, \binits{M.}},
\bauthor{\bsnm{Piccioli}, \binits{F.S.}},
\bauthor{\bsnm{Ristanović}, \binits{Z.}},
\bauthor{\bsnm{Lombardi}, \binits{P.}},
\bauthor{\bsnm{Moradi}, \binits{A.}},
\bauthor{\bsnm{Adhikari}, \binits{S.}},
\bauthor{\bsnm{Deperasinska}, \binits{I.}},
\bauthor{\bsnm{Kozankiewicz}, \binits{B.}},
\bauthor{\bsnm{Orrit}, \binits{M.}},
\bauthor{\bsnm{Toninelli}, \binits{C.}}:
\batitle{Laser-induced frequency tuning of fourier-limited single-molecule
  emitters}.
\bjtitle{ACS Nano}
\bvolume{14}(\bissue{10}),
\bfpage{13584}--\blpage{13592}
(\byear{2020})
\doiurl{10.1021/acsnano.0c05620} .
\bcomment{PMID: 32936612}
\end{barticle}
\endbibitem

\bibitem[\protect\citeauthoryear{Pazzagli et~al.}{2018}]{pazzagli2018self}
\begin{barticle}
\bauthor{\bsnm{Pazzagli}, \binits{S.}},
\bauthor{\bsnm{Lombardi}, \binits{P.}},
\bauthor{\bsnm{Martella}, \binits{D.}},
\bauthor{\bsnm{Colautti}, \binits{M.}},
\bauthor{\bsnm{Tiribilli}, \binits{B.}},
\bauthor{\bsnm{Cataliotti}, \binits{F.S.}},
\bauthor{\bsnm{Toninelli}, \binits{C.}}:
\batitle{Self-assembled nanocrystals of polycyclic aromatic hydrocarbons show
  photostable single-photon emission}.
\bjtitle{ACS Nano}
\bvolume{12}(\bissue{5}),
\bfpage{4295}--\blpage{4303}
(\byear{2018})
\end{barticle}
\endbibitem

\bibitem[\protect\citeauthoryear{Nicolet et~al.}{2007}]{nicolet2007single}
\begin{barticle}
\bauthor{\bsnm{Nicolet}, \binits{A.A.L.}},
\bauthor{\bsnm{Bordat}, \binits{P.}},
\bauthor{\bsnm{Hofmann}, \binits{C.}},
\bauthor{\bsnm{Kol'chenko}, \binits{M.A.}},
\bauthor{\bsnm{Kozankiewicz}, \binits{B.}},
\bauthor{\bsnm{Brown}, \binits{R.}},
\bauthor{\bsnm{Orrit}, \binits{M.}}:
\batitle{Single dibenzoterrylene molecules in an anthracene crystal: Main
  insertion sites}.
\bjtitle{ChemPhysChem}
\bvolume{8}(\bissue{13}),
\bfpage{1929}--\blpage{1936}
(\byear{2007})
\doiurl{10.1002/cphc.200700340}
\end{barticle}
\endbibitem

\bibitem[\protect\citeauthoryear{Rezai et~al.}{2018}]{rezai2018coherence}
\begin{barticle}
\bauthor{\bsnm{Rezai}, \binits{M.}},
\bauthor{\bsnm{Wrachtrup}, \binits{J.}},
\bauthor{\bsnm{Gerhardt}, \binits{I.}}:
\batitle{Coherence properties of molecular single photons for quantum
  networks}.
\bjtitle{Phys. Rev. X}
\bvolume{8},
\bfpage{031026}
(\byear{2018})
\doiurl{10.1103/PhysRevX.8.031026}
\end{barticle}
\endbibitem

\bibitem[\protect\citeauthoryear{Duquennoy
  et~al.}{2023}]{duquennoy2023singular}
\begin{barticle}
\bauthor{\bsnm{Duquennoy}, \binits{R.}},
\bauthor{\bsnm{Colautti}, \binits{M.}},
\bauthor{\bsnm{Lombardi}, \binits{P.}},
\bauthor{\bsnm{Berardi}, \binits{V.}},
\bauthor{\bsnm{Gianani}, \binits{I.}},
\bauthor{\bsnm{Toninelli}, \binits{C.}},
\bauthor{\bsnm{Barbieri}, \binits{M.}}:
\batitle{Singular spectrum analysis of two-photon interference from distinct
  quantum emitters}.
\bjtitle{Phys. Rev. Res.}
\bvolume{5},
\bfpage{023191}
(\byear{2023})
\doiurl{10.1103/PhysRevResearch.5.023191}
\end{barticle}
\endbibitem

\bibitem[\protect\citeauthoryear{Wrigge et~al.}{2008}]{wrigge2008efficient}
\begin{barticle}
\bauthor{\bsnm{Wrigge}, \binits{G.}},
\bauthor{\bsnm{Gerhardt}, \binits{I.}},
\bauthor{\bsnm{Hwang}, \binits{J.}},
\bauthor{\bsnm{Zumofen}, \binits{G.}},
\bauthor{\bsnm{Sandoghdar}, \binits{V.}}:
\batitle{Efficient coupling of photons to a single molecule and the observation
  of its resonance fluorescence}.
\bjtitle{Nat. Phys.}
\bvolume{4}(\bissue{1}),
\bfpage{60}--\blpage{66}
(\byear{2008})
\doiurl{10.1038/nphys812}
\end{barticle}
\endbibitem

\bibitem[\protect\citeauthoryear{Hail et~al.}{2019}]{hail2019nanoprinting}
\begin{barticle}
\bauthor{\bsnm{Hail}, \binits{C.U.}},
\bauthor{\bsnm{H{\"o}ller}, \binits{C.}},
\bauthor{\bsnm{Matsuzaki}, \binits{K.}},
\bauthor{\bsnm{Rohner}, \binits{P.}},
\bauthor{\bsnm{Renger}, \binits{J.}},
\bauthor{\bsnm{Sandoghdar}, \binits{V.}},
\bauthor{\bsnm{Poulikakos}, \binits{D.}},
\bauthor{\bsnm{Eghlidi}, \binits{H.}}:
\batitle{Nanoprinting organic molecules at the quantum level}.
\bjtitle{Nat. Commun.}
\bvolume{10}(\bissue{1}),
\bfpage{1880}
(\byear{2019})
\doiurl{10.1038/s41467-019-09877-5}
\end{barticle}
\endbibitem

\bibitem[\protect\citeauthoryear{Rattenbacher
  et~al.}{2019}]{rattenbacher2019coherent}
\begin{barticle}
\bauthor{\bsnm{Rattenbacher}, \binits{D.}},
\bauthor{\bsnm{Shkarin}, \binits{A.}},
\bauthor{\bsnm{Renger}, \binits{J.}},
\bauthor{\bsnm{Utikal}, \binits{T.}},
\bauthor{\bsnm{G{\"o}tzinger}, \binits{S.}},
\bauthor{\bsnm{Sandoghdar}, \binits{V.}}:
\batitle{Coherent coupling of single molecules to on-chip ring resonators}.
\bjtitle{New J. Phys.}
\bvolume{21}(\bissue{6}),
\bfpage{062002}
(\byear{2019})
\doiurl{10.1088/1367-2630/ab28b2}
\end{barticle}
\endbibitem

\bibitem[\protect\citeauthoryear{Gurlek et~al.}{2021}]{gurlek2021engineering}
\begin{barticle}
\bauthor{\bsnm{Gurlek}, \binits{B.}},
\bauthor{\bsnm{Sandoghdar}, \binits{V.}},
\bauthor{\bsnm{Martin-Cano}, \binits{D.}}:
\batitle{Engineering long-lived vibrational states for an organic molecule}.
\bjtitle{Phys. Rev. Lett.}
\bvolume{127},
\bfpage{123603}
(\byear{2021})
\doiurl{10.1103/PhysRevLett.127.123603}
\end{barticle}
\endbibitem

\bibitem[\protect\citeauthoryear{Zirkelbach et~al.}{2022}]{zirkelbach2022high}
\begin{barticle}
\bauthor{\bsnm{Zirkelbach}, \binits{J.}},
\bauthor{\bsnm{Mirzaei}, \binits{M.}},
\bauthor{\bsnm{Deperasińska}, \binits{I.}},
\bauthor{\bsnm{Kozankiewicz}, \binits{B.}},
\bauthor{\bsnm{Gurlek}, \binits{B.}},
\bauthor{\bsnm{Shkarin}, \binits{A.}},
\bauthor{\bsnm{Utikal}, \binits{T.}},
\bauthor{\bsnm{Götzinger}, \binits{S.}},
\bauthor{\bsnm{Sandoghdar}, \binits{V.}}:
\batitle{{High-resolution vibronic spectroscopy of a single molecule embedded
  in a crystal}}.
\bjtitle{J. Chem. Phys.}
\bvolume{156}(\bissue{10}),
\bfpage{104301}
(\byear{2022})
\doiurl{10.1063/5.0081297}
\end{barticle}
\endbibitem

\end{thebibliography}

\end{document}